\documentclass[onecolumn,secnumarabic,amssymb,nofootinbib,notitlepage,aps,pra]{revtex4-1}

\usepackage[utf8x]{inputenc}
\usepackage{amsmath}
\usepackage[a4paper,vmargin={20mm,20mm}]{geometry}
\usepackage{bbold}
\usepackage{amsthm}
\usepackage{color}
\usepackage{adjustbox}
\usepackage{graphicx}
\usepackage{tabularx}

\usepackage{amssymb}

\usepackage{rotating}

\newcommand{\ket}[1]{\left| #1\right. \rangle}

\newcommand{\dd}{\mathrm{d}}

\newtheorem{thm}{Theorem}
\newtheorem*{thm*}{Theorem}
\newtheorem{crl}{Corollary}

\begin{document}

\title{Measurement dependent locality}
\author{Gilles P\"utz}
\affiliation{University of Geneva}
\affiliation{ETH Zurich}
\author{Nicolas Gisin}
\affiliation{University of Geneva}
\date{\today}

\begin{abstract}
The demonstration and use of Bell-nonlocality, a concept that is fundamentally striking and is at the core of applications in device independent quantum information processing, relies heavily on the assumption of measurement independence, also called the assumption of free choice. The latter cannot be verified or guaranteed. In this paper, we consider a relaxation of the measurement independence assumption. We briefly review the results of \cite{Putz14}, which show that with our relaxation, the set of so-called measurement dependent local (MDL) correlations is a polytope, i.e. it can be fully described using a finite set of linear inequalities. Here we analyze this polytope, first in the simplest case of 2 parties with binary inputs and outputs, for which we give a full characterization. We show that partially entangled states are preferable to the maximally entangled state when dealing with measurement dependence in this scenario. We further present a method which transforms any Bell-inequality into an MDL inequality and give valid inequalities for the case of arbitrary number of parties as well as one for arbitrary number of inputs. We introduce the assumption of independent sources in the measurement dependence scenario and give a full analysis for the bipartite scenario with binary inputs and outputs. Finally, we establish a link between measurement dependence and another strong hindrance in certifying nonlocal correlations: nondetection events.
\end{abstract}

\maketitle
\section{Introduction}

For decades after the advent of quantum mechanics, the question of whether hidden parameters could explain the correlations between entangled particles remained open. John Bell~\cite{Bell1964} finally answered the question in the negative: no local explanation for the correlations can hold. Explaining this fact to physicists even nowadays leads to a surprised raising of the eyebrows as the concept of nonlocality at least at first seems very counterintuitive, and many will try to argue one way or the other that this cannot be true. And while this fact alone makes nonlocality a highly interesting subject to study for our understanding of fundamental physics, it becomes even more important after noticing that it can be used in modern day applications. The fact that nonlocal correlations have properties like inherent randomness~\cite{Colbeck2006} and monogamy of correlations gave rise to the field of device independent quantum information processing~\cite{Barrett05,Brunner:RMP}: certain tasks in information processing can be accomplished with minimal assumptions on the devices used. In fact the user of a device may not need to trust or understand the inner workings of the device itself and still guarantee that the task is completed successfully purely based on the observed correlations. This has found uses for example in cryptography, specifically quantum key distribution~\cite{Ekert91,BarrettKent05,Acin06}, randomness generation, specifically randomness expansion~\cite{Pironio2010} and amplification~\cite{Colbeck2012} and other quantum information processing tasks, like entanglement certification\cite{Bancal11,Barreiro13}.\\

Let us start by restating the precise formulation of locality. Consider the following scenario: two parties, Alice and Bob, separate in their respective labs, each receive a particle from a source situated between them. Then, each of them chooses a measurement from a list of possible measurements and perform this measurement on the particle and record the outcome. They will not use any information about the state of the particle or the precise measurements, instead they just record the labels of the measurements, henceforth called the input, as well as the respective outcomes. We denote by $x$ and $y$ the input and by $a$ and $b$ the outcome of Alice and Bob, respectively. Furthermore, we denote by $\lambda$ the possible common pasts that both particles share\footnote{The mathematically precise notation will be introduced in section \ref{technicalities}.}. Using nothing but the properties of probability distributions, see section \ref{secprobabilities}, we have that
\begin{align}
p(abxy)=\int\dd\lambda\rho(\lambda)p(xy|\lambda)p(ab|xy\lambda).
\end{align}

We can now introduce the three assumptions that comprise the concept of locality:
\begin{itemize}
\item \textit{Outcome independence:} Any correlation between the outcomes comes from their inputs as well as their common past: \\
$p(ab|xy\lambda)=p(a|xy\lambda)p(b|xy\lambda)$
\item \textit{Parameter independence:} The outcome on one side cannot depend on the input of the other side:\\
$p(a|xy\lambda)=p(a|x\lambda)$ and $p(b|xy\lambda)=p(b|y\lambda)$
\item \textit{Measurement independence:} The inputs were chosen independently of the common past of the particles:\\
$p(xy|\lambda)=p(xy)$
\end{itemize}

Together, and dividing both sides by $p(xy)$, we find the usual definition of local correlations
\begin{align}
\label{local}
p(ab|xy)=\int\dd\lambda\rho(\lambda)p(a|x\lambda)p(b|y\lambda).
\end{align}
While we do not go into the details of the proof here, it is important to note that for a fixed number of inputs and outputs on each side, these correlations form a convex polytope, a convex geometric structure with a finite set of vertices. A convex polytope has the useful property that it can, equivalently, be described by a set of linear inequalities instead of by its vertices. Within the context of local probability distributions, these inequalities are called Bell inequalities. The most well-known example of such an inequality is given in the scenario where Alice and Bob choose from 2 possible inputs ($0$ and $1$) and receive binary outcomes: the CHSH Bell inequality\cite{Clauser1969}. We rewrite it here in the form, equivalent under nosignaling, first introduced by Eberhard~\cite{Eberhard93}
\begin{align}
\label{Eberhard}
p(00|00)-p(01|01)-p(10|10)-p(00|11)\leq 0.
\end{align}
Any local distribution, meaning any probability distribution that can be written in the form (\ref{local}) satisfies this inequality.\\

The interesting part is that there are correlations which can be realized by quantum mechanics that violate this inequality, which entails certain interesting properties independently of how exactly the correlations were realized. These properties are based on the fact that no common past can explain the correlations. It is therefore important to ensure that if such a common past were to exist, it would have to satisfy the outcome independence, parameter independence and measurement independence assumptions. When it comes to outcome and parameter independence, this is usually not seen as a problem: if we assume that information transfer needs a physical carrier, then we can in principle enforce outcome and parameter independence by performing the experiment in spacelike separation. Failing that, at least when it comes to applications it is usually still reasonable to assume that no signal is transmitted between the devices of Alice and Bob. The reason for this is that the tasks that nonlocal correlations are useful for are mostly linked to some form of privacy, like cryptography. A transmitter in the device would render such a task inherently impossible. Overall, it is fair to say that the parameter independence assumption is well-motivated . It is assumed to hold for the rest of this work.

The same can however not be said for the assumption of measurement independence. From a foundational perspective, a total independence of the input-choice $x$ and $y$ and the common past $\lambda$ of the particles may seem acceptable, but is still an uncomfortable assumption that cannot be enforced. When it comes to applications it gets even worse: if we consider the common past $\lambda$ to represent the influence of some adversary in a privacy related information theoretic task, then the measurement independence assumption means that for some reason this adversary has no way at all to influence the random number generators used to generate the inputs. We have to assume that these are fully safe and unpredictable. Weakening this assumption is the focus of this work.

\begin{figure}
\label{figmdlscenario}
\centering
\includegraphics[width=0.5\textwidth]{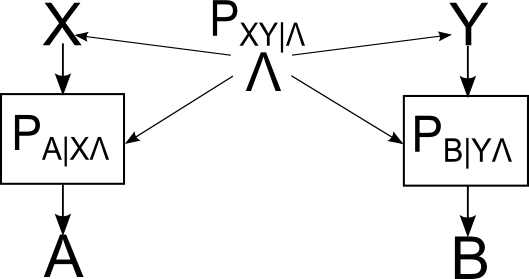}
\caption{A bipartite scenario with measurement dependent local hidden variables. A source emits particles carrying the common information $\Lambda$ towards the experimenters. The latter then choose measurements labeled by $X$ and $Y$ respectively and get outcomes $A$ and $B$. The choise of measurements is correlated to the particles common past $\Lambda$.}
\end{figure}

Unfortunately, it is not possible to fully eliminate the measurement independence assumption. Without it, outcome independent and parameter independent distributions can reproduce all quantum correlations and they become useless within this framework. We can however weaken the assumption as much as possible by replacing it with a less restrictive form, see figure \ref{figmdlscenario}. Such a weakening has been studied before in different works~\cite{Hall2011,Barrett2011,Thinh2013}, but here we will take a slightly different approach. Specifically, for fixed parameters $\ell$ and $h$ with $[\ell,h]\subsetneqq [0,1]$, we replace assumption 3 by
\begin{align}
\label{md}
\exists\ell,h\text{ s.t. }\ell\leq p(xy|\lambda)\leq h\text{ }\forall x,y,\lambda.
\end{align}
In other words we do allow the common past $\lambda$ to influence the inputs, but only to the degree that each inputpair $(x,y)$ has still at least a probability of $\ell$ and at most a probability of $h$ to appear. This particular constraint measures the amount of randomness in the inputs by limiting their deviation from uniform even when conditioned on the common past $\lambda$. It corresponds to the single-shot version of the Santha-Vazirani condition
\begin{align}
\epsilon\leq p(x_i|x1\ldots x_{i-1}\lambda)\leq 1-\epsilon,
\end{align}
which is the constraint usually considered for the cases of randomness amplification~\cite{Colbeck2012}. We thus study the correlations that are of the form
\begin{align}
\label{mdl}
p(abxy)=\int\dd\lambda\rho(\lambda)p(xy|\lambda)p(a|x\lambda)p(b|y\lambda)
\end{align}
and fulfill assumption (\ref{md}). We call these correlations \textit{measurement dependent local}. We introduced them originally in \cite{Putz14} where we showed that they form, like the local correlations, a polytope. We also demonstrated that, maybe surprisingly, there are quantum correlations that cannot be reproduced by such measurement dependent local correlations as long as $\ell>0$. This means that, even though the measurement independence assumption cannot simply be removed, it can be made arbitrarily weak.

In the rest of the present work, we study these measurement dependent local correlations in different settings. We first clarify notations and definitions and state the technical results from which our results follow in section \ref{technicalities}. We then briefly recapitulate the results of \cite{Putz14}, before turning to new results starting with a more extensive study of the scenario with 2 parties with binary inputs and outputs. In section \ref{secgeneral}, we show that any Bell-inequality can be transformed into a measurement dependent local inequality. We use this result to study the scenario of $N$ parties with binary inputs and outputs. The generalised scenario of 2 parties with arbitrary inputs and outputs is also considered, and we show that even $\ell=0$ still does not allow to reproduce all quantum mechanical correlations. We introduce the additional reasonable assumption of independent sources and study the resulting correlations in section \ref{secindsources}. Finally, we introduce the concept of limited detection locality and show how detection inefficiencies can be linked to measurement dependence in section \ref{LDL}.

\section{Technicalities}
\label{technicalities}
In this section we introduce the mathematical concepts used in this paper, clarify notations and definitions and state the technical theorems from which several of results follow. 

\subsection{Polytopes}
\label{polytopes}
\begin{figure}
\label{figpolytope}
\centering
\includegraphics[width=0.5\textwidth]{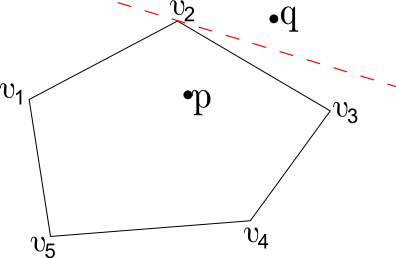}
\caption{A polytope in 2 dimensions. The set of vertices is given by $\mathcal{V}=\{v_i\}_{i=1}^5$ and the facets are given by the solid black lines. In 2 dimensions, every facet is given by 2 vertices. This is however not the case in general dimensions. To determine whether a point lies within the polytope, we check that it lies on the right side of all the facets, as it is the case with the point $p$. The point $q$ lies on the wrong side of the facet $(v_2v_3)$ and is thus not part of the polytope. It is often computationally difficult to find all the facets. Hyperplanes like the red dashed line can in such cases still be useful, as it would still allow one to determine that the point $q$ is not a member of the polytope.}
\end{figure}

A polytope is a convex geometric structure, which we write as a set of vectors, that has a finite number of extremal points, called vertices. Since it is convex it is fully described by the set of extremal points, all other points that are part of the structure are given by convex combinations of the vertices. Equivalently, a polytope can be described by a finite set of hyperplanes which are such that all the points inside the polytope lie on one side of each hyperplane. These are called the facets of the polytope. See figure \ref{figpolytope} for an example in 2 dimensions.\\

\textit{Vertices:} Given a polytope $\mathcal{P}$, we denote by $\mathcal{V_P}=\{v_\mathcal{P}^i\}_{i=1}^n$ the set of its vertices. Every point inside the polytope can be written as a convex combination of these vertices, $p\in\mathcal{P}\Leftrightarrow p=\sum_i\alpha_i v_\mathcal{P}^i$ with $\alpha_i\geq 0$, $\sum_i\alpha_i=1$. Besides being able to generate every point of the polytope, the set of vertices can be useful when optimizing convex quantities: any convex function from elements of the polytope to $\mathbb{R}$ will take its global maximum at at least one of the vertices. The same holds for the global minimum.\\

\textit{Facets:} We denote by $\mathcal{F_P}=\{(f_i,B_i)\}_{i=1}^m$ the set of facets of the polytope. A facet is a hyperplane and thus described by a linear equality of the form $f\cdot p= B$,  where $\cdot$ denotes the scalar product. All the points of the polytope lie on the same side of each facet, i.e. $p\in\mathcal{P}\Rightarrow f\cdot p\leq B$\footnote{Technically, we could also have $p\in\mathcal{P}\Rightarrow f\cdot p\geq B$. In this case we simply set $f\rightarrow -f$ and $B\rightarrow -B$ since this does not change the equation of the hyperplane $f\cdot p=B$. We can thus choose $\leq$ by convention.} Additionally, if a point lies on the correct side of all the facets, then it follows that it is a member of the polytope, such that we have $f\cdot p\leq B$ $\forall (f,B)\in\mathcal{F_P}\Rightarrow p\in\mathcal{P}$. The facets are useful to determine whether a point lies inside the polytope or not since one only needs to check whether or not it respects all the inequalities given by $\mathcal{F_P}$.\\

It is possible to find $\mathcal{F_P}$ given $\mathcal{V_P}$ and vice versa. We will not elaborate on how this is done here, it suffices to say that if can be done using programs like, for example, the porta-library~\cite{Porta}. Unfortunately, it is computationally expensive to perform this transformation. It is therefore useful to establish inequalities that, even though they are not facets, hold for all the elements of the polytope. A violation of such an inequality then implies that a point is not a member of the polytope. Within the context of nonlocality, randomness generation and similar, we are usually interested in certifying precisely such non-memberships.\\

In \cite{Putz14}, we proved a general theorem on polytopes: Given two polytopes, they can be combined in a specific way such that the resulting structure forms again a polytope. If the vertices of the two composing polytopes are known, then so are the vertices of the resulting polytope. The measurement dependent local correlations we study in this paper form such a polytope.

\begin{thm}
\label{polytopethm}
Let $\mathcal{P}\subset\mathbb{R}^{m'n}$ and $\mathcal{Q}\subset\mathbb{R}^{mn}$ be two polytopes. Let the components of the vectors of $\mathcal{P}$ be labelled by $p(k,l)$ and the ones of $\mathcal{Q}$ by $q(k,l')$ where $k=1\ldots n$, $l=1\ldots m$ and $l'=1\ldots m'$. Let
\begin{align}
\mathcal{R}=\big\{r\in\mathbb{R}^{mm'n} : &\exists\Lambda\text{ measureable set},\\
&\exists\rho:\Lambda\rightarrow [0,1]\text{ with }\int_\Lambda\dd\lambda\rho(\lambda)=1\nonumber\\
&\exists \{p_\lambda\}_{\lambda\in\Lambda}\subset\mathcal{P}, \quad\exists \{q_{\lambda}\}_{\lambda\in\Lambda}\subset\mathcal{Q} \text{ s.t.}\nonumber\\
&r(k,l,l')=\int\dd\lambda\rho(\lambda)p_\lambda(k,l)q_{\lambda}(k,l')\quad \forall k\in\{1\ldots n\},l\in\{1\ldots m\},l'\in\{1\ldots m'\}\big\}.\nonumber
\end{align}
Let $\mathcal{V_P}$ and $\mathcal{V_Q}$ be the sets of vertices of $\mathcal{P}$ and $\mathcal{Q}$ respectively.\\
Then $\mathcal{R}$ is a polytope and its set of vertices $\mathcal{V_R}$ fulfils
\begin{align}
\mathcal{V_R}\subseteq\big\{v_\mathcal{R}^{ij} : v_\mathcal{R}^{ij}(k,l,l')=v_\mathcal{P}^i(k,l)v_\mathcal{Q}^j(k,l')\quad i\in\{1\ldots |\mathcal{V_P}|\},j\in\{1\ldots |\mathcal{V_Q}|\}\big\}.
\end{align}
\end{thm}

The proof of this theorem can be found in the appendix. Note that it is unproblematic that the theorem only shows that the vertices of $\mathcal{R}$ are a subset of the set of combined vertices; all of the points in the set of combined vertices, of which there are a finite amount, lie inside the polytope, they may just not be vertices themselves. These nonextremal points can easily be eliminated by checking for each of the elements of the set of combined vertices whether it can be written as a convex combination of the others.\\

\subsection{Random variables and probabilities}
\label{secprobabilities}
As already apparent in the introduction, this work relies heavily on probability distributions. We therefore define them here.

\textit{Probability distribution or density:} A probability distribution is a function $p$ from some discrete set $S$ into $[0,1]$ such that
\begin{align}
\label{positivity}
p(s)&\geq 0\quad\forall s\in S\\
\label{normalization}
\sum_{s\in S}p(s)&=1.
\end{align}
Analogously, a probability density is a function $\rho$ from a non-discrete measurable set $C$ into $[0,1]$ such that
\begin{align}
\rho(c)&\geq 0\quad\forall c\in C\\
\int_C\dd c\rho(c) &=1.
\end{align}
Throughout the paper, we will denote distributions over a discrete set by $p$ and densities over a (potentially) non-discrete set by $\rho$.

\textit{Random variable:} A random variable $V$ is a variable taking values in a given measurable set, called the alphabet, with an associated probability distribution. In the following, we use capital letters to denote random variables and the corresponding lower case letters to denote the values of the variables. The probability distribution over the random variable $V$ is denoted by $p_V$ and the probability of $V$ taking the value $v$ by $p_V(v)$. To ease notation, we will often just write $p(v)$ if the random variable is clear. The same goes for the case of continuous measurable alphabets and probability densities.\\

\textit{Joint distribution:} Given two random variables $V$ and $W$, a joint probability distribution $p_{VW}$ may be defined. It satisfies
\begin{align}
p_V(v)=\sum_{w}p_{VW}(vw)
\end{align}
and similarly for $W$. $p_V$ and $p_W$ are called the marginals of $p_{VW}$.\\

\textit{Conditional distribution:} Given two random variables $V$ and $W$ with a joint distribution $p_{VW}$, we define the probability distribution over $V$ conditioned on $W$ by
\begin{align}
p_{V|W}(v|w)=\frac{p_{VW}(vw)}{p_W(w)}.
\end{align}
This is known as Bayes' rule. $p_{V|W}(v|w)$ represents the probability of $V$ taking value $v$ given the knowledge that $W$ took value $w$.

\subsection{Locality, nonsignaling and measurement dependence}
\label{lnsmdl}
Let us now get back to the topic of this paper. We consider a fixed number of distinct parties between which particles are distributed from a common source. The parties then perform different possible measurements, labelled by an input, on these particles and receive an outcome each.\\

\textit{Scenario:} A scenario is defined by a vector of integers $S=\big(N,\{n_i\}_{i=1}^N,\{\{m_i^j\}_{j=0}^{n_i-1}\}_{i=1}^N\big)$, which represent, in order, the number of parties in the scenario, the number of inputs for each party and the number of outcomes for each input of each party. Very often we consider simplified scenarios where, for example, each party has the same number of inputs or each input has the same number of outputs. We use a simplified notation in this case. For example the notation $(N,n,m)$ denotes the scenario of $N$ parties, each having $n$ inputs and $m$ outcomes for each input.\\

\textit{Random variables of the scenario:} We denote by $X_i$ the random variable modelling the input of the $i$-th party, with alphabet $\{0\cdots n_i-1\}$. For the outputs of the $i$-th party, we define the random variable $A_i$ with alphabet $\{0\cdots\max_j m_i^j-1\}$ with the property that the corresponding probability distribution satisfies $p_{A_iX_i}(ax)=0$ if $a\geq m_i^x$. This last condition is due to the fact that for input $x$ only $m_i^x$ different outcomes can occur. Additionally, we denote by $\Lambda$ the random variable describing the common past of the particles. We do not restrict its alphabet.\\

In the following, we define interesting sets of probability distributions. Since these sets will all contain probability distributions only, we do not specify it again in the definition of the sets.\\

\textit{Locality:} Local correlations have been presented in the introduction. They arise from the assumptions of outcome independence, parameter independence and measurement independence. The set of local correlations for a fixed scenario $S$ is defined by
\begin{align}
\label{localpolytope}
\mathcal{L}_{S}=\big\{p_{\{A_i\}_{i=1}^N|\{X_i\}_{i=1}^N} : &\exists \Lambda \text{ random variable and }\rho_{\{A_i\}_{i=1}^N\Lambda|\{X_i\}_{i=1}^N}\\
&p(\{a_i\}_{i=1}^N|\{x_i\}_{i=1}^N)=\int_\Lambda\dd\lambda\rho(\lambda)\prod_{i=1}^Np(a_i|x_i\lambda)\big\}.\nonumber
\end{align}
Whenever the scenario $S$ is clear, we will omit it to ease notation. It is a well-known fact that the set of local correlations for a fixed scenario forms a polytope, the vertices of which are given by
\begin{align}
\mathcal{V_L}=\big\{v_{\{A_i\}_{i=1}^N|\{X_i\}_{i=1}^N} : &\forall i\in\{1\ldots N\}\forall x_i\in\{0\ldots n_i-1\} \exists! a_i^{x_i}\text{ s.t. }\\
&v(\{a_i\}_{i=1}^N|\{x_i\}_{i=1}^N)=\prod_i\delta_{a_i=a_i^{x_i}}\big\}.\nonumber
\end{align}
These are called the determinstic points since the corresponding probability distributions are deterministic, i.e. only take values $0$ or $1$. It is worth to note that this could also have been proven using our polytope-theorem \ref{polytopethm} iteratively. \\

\textit{Nosignaling:} Nonsignaling distributions have the properties of parameter independence and measurement independence, but do not satisfy outcome independence. The name comes from the fact that this is the maximal set of distributions that cannot be used to transmit a signal among the parties, i.e. the outcome of party $i$ does not reveal any information about the inputs of the other parties. For a given scenario $S$, that again will usually be omitted in the rest of the paper, we define the set of nonsignaling distributions by
\begin{align}
\label{nspolytope}
\mathcal{NS}_{S}=\big\{p_{\{A_i\}_{i=1}^N|\{X_i\}_{i=1}^N} : p(a_j|\{x_i\}_{i=1}^N)=p(a_j|x_j)\quad\forall j,a_j,\{x_i\}_{i=1}^N\big\}.
\end{align}
This set forms again a polytope, which can be seen by the fact that it is only subject to linear constraints in the form of the positivity and normalization requirement of probability distributions as well as the nonsignaling condition $p(a_j|\{x_i\}_{i=1}^N\lambda)=p(a_j|x_j\lambda)$. The set of all quantum correlations is nonsignaling and therefore included in the nonsignaling polytope. One of the main reasons to study nonsignaling correlations is due to the fact that the set of quantum correlations is difficult to characterize compared to the nonsignaling polytope.\\

\textit{Measurement dependent locality:} The idea behind measurement dependent local distributions has already been introduced in the introduction. Formally, we consider a fixed scenario $S=\big(N,\{n_i\}_{i=1}^N,\{\{m_i^j\}_{j=1}^{n_i}\}_{i=1}^N\big)$ as well as fixed parameters $\ell$ and $h$, which will usually be omitted from notation in the following, with
\begin{align}
\label{lhbounds}
\max(1-\big((\prod_{i=1}^Nn_i)-1\big)h,0)\leq\ell\leq \frac{1}{\prod_{i=1}^Nn_i}\leq h\leq 1-\big((\prod_{i=1}^Nn_i)-1\big)\ell.
\end{align}
The set of measurement dependent local correlations for these $\ell$ and $h$ is then given by
\begin{align}
\label{mdlpolytope}
\mathcal{MDL}_S(\ell,h)=\big\{p_{\{A_i\}_{i=1}^N\{X_i\}_{i=1}^N} : &\exists \Lambda \text{ random variable and }\rho_{\{A_i\}_{i=1}^N\Lambda\{X_i\}_{i=1}^N}\\
&p(\{a_i\}_{i=1}^N\{x_i\}_{i=1}^N)=\int_\Lambda\dd\lambda\rho(\lambda)p(\{x_i\}_{i=1}^N|\lambda)\prod_{i=1}^Np(a_i|x_i\lambda)\nonumber\\
&\ell\leq p(\{x_i\}_{i=1}^N|\lambda)\leq h\quad\forall \{x_i\}_{i=1}^N,\lambda\big\}.\nonumber
\end{align}
The bounds on $\ell,h$ given by (\ref{lhbounds}) follow from the fact that probability distributions are positive (\ref{positivity}) and normalised (\ref{normalization}), making parameters $(\ell,h)$ that do not respect the denoted bounds either trivially satisfied (e.g. $\ell< 0$) or impossible to satisfy (e.g. $\ell> \frac{1}{\prod_{i=1}^Nn_i}$). The set of input-distributions for a given $\lambda$, i.e. the set of the possible $p(\{x_i\}_{i=1}^N|\lambda)$ is given by
\begin{align}
\mathcal{I}(\ell,h)=\big\{ p_{\{X_i\}_{i=1}^N} : \ell\leq p(\{x_i\}_{i=1}^N)\leq h\quad \forall \{x_i\}_{i=1}^N\big\}.\nonumber
\end{align}
Since the constraints are all linear, this is a polytope. The vertices are the points that saturate as many of the inequalities as possible\footnote{Any point that does not can be written as a convex combination of two points which saturate all the inequalities that the point does as well as at least one more.}. Defining $t=\lfloor\frac{(\prod_in_i)h-1}{h-l}\rfloor$ the vertices are therefore given by
\begin{align}
\label{Vi}
\mathcal{V_I}(\ell,h)=\big\{v_{\{X_i\}_{i=1}^N} :& \exists\pi\text{ permutation s.t. } \\
&v=\pi\big((\underbrace{\ell\ldots\ell}_{t},\underbrace{h\ldots h}_{(\prod_in_i) -t-1},1-t\ell-((\prod_in_i) -t-1)h)\big)\big\}.\nonumber
\end{align}

Our analysis is made possible by the fact that the measurement dependent local distributions, just like the local and nonsignaling distributions, form a polytope.
\begin{crl}
\label{mdlthm}
For fixed $S$, $\ell$ and $h$, $\mathcal{MDL}_S(\ell,h)$ forms a polytope. Its vertices are given by
\begin{align}
\label{Vmdl}
\mathcal{V_{MDL}}(\ell,h)\subseteq\big\{v^{ij}_{\{A_i\}_{i=1}^N,\{X_i\}_{i=1}^N} : &v^{ij}(\{a_i\}_{i=1}^N,\{x_i\}_{i=1}^N)=v^{i}_\mathcal{I}(\{x_i\}_{i=1}^N)v^j_\mathcal{L}(\{a_i\}_{i=1}^N|\{x_i\}_{i=1}^N)\nonumber\\
&v^{i}_\mathcal{I}(\{x_i\}_{i=1}^N)\in \mathcal{V_I}(\ell,h),\quad v^{j}_\mathcal{L}\in \mathcal{V_L}\big\}_{i,j}.
\end{align}
\end{crl}
This is a direct corollary of the polytope-theorem \ref{polytopethm}, since the measurement dependent set is obtained by combining the input polytope $\mathcal{I}(\ell,h)$ with the local polytope $\mathcal{L}$.

It is interesting to note that the MDL-polytope is indeed not a subset of the nonsignaling polytope. The fact that the hidden variable is correlated with the input of one party and the outcome of the others permits to establish correlations between them, in other words it permits to signal.

\section{The (2,2,2) scenario}
\label{222}
We consider here the simplest case of $2$ parties with binary inputs and outputs on each side. To simplify notation, we call these parties Alice and Bob and label their inputs by $X$ and $Y$ and their outputs by $A$ and $B$ respectively. For fixed $\ell$ and $h$, we want to know whether a given correlation could be explained by measurement dependent local (MDL) resources, i.e. whether it is of the form
\begin{align}
\label{mdlcorr}
p(abxy)=\int\dd\lambda\rho(\lambda)p(xy|\lambda)p(a|x\lambda)p(b|y\lambda)\\
\ell\leq p(xy|\lambda)\leq h\quad \forall x,y,\lambda\nonumber
\end{align}
for fixed $\ell$ and $h$ with $[\ell,h]\subsetneqq [0,1]$. In section \ref{technicalities}, we fully characterized the set of MDL correlations by showing that they can all be written as a convex combination of a finite set of vertices, i.e. they form a polytope. Since we are mostly interested in determining membership, finding inequalities respected by all MDL correlations is of interest.\\

Let us first recapitulate the results of our previous paper~\cite{Putz14}. Using the polytope structure, we showed that all MDL correlations satisfy the inequality
\begin{align}
\label{goldenineq}
\ell p(0000)-h\big(p(0101)+p(1010)+p(0011)\big)\leq 0.
\end{align}
This inequality turns out to be very useful when looking at quantum correlations. In fact, maybe surprisingly, there exist quantum correlations which can violate this inequality $\forall\ell>0$ and $\forall h$. To this end, we need to find quantum correlations that satisfy
\begin{align*}
p(0000)>0,\quad p(0101)=p(1010)=p(0011)=0.
\end{align*}
This is a formulation of Hardy's paradox~\cite{Hardy93}, which was shown to be realizable using any partially entangled pure 2-qubit state. For example, it can be realized using the 2-qubit state
\begin{align}
\label{goldenstate}
\ket{\Psi}=\frac{1}{\sqrt{3}}\big(\ket{01}+\ket{10}-\ket{11}\big)
\end{align}
if Alice and Bob both measure in the basis $\left\{\frac{\ket{0}+\ket{1}}{\sqrt{2}},\frac{\ket{0}-\ket{1}}{\sqrt{2}}\right\}$ for input $0$ and in the basis $\left\{\ket{0},\ket{1}\right\}$ for input $1$. This state was implemented experimentally by our collaborators~\cite{Aktas2015}\footnote{Note that in the paper the state (\ref{goldenstate}) is written in its Schmidt basis.}, which due to noise limitations violated inequality (\ref{goldenineq}) $\forall\ell>0.090$.\\

Let us stress this result: Measurement independence was a crucial assumption in deriving nonlocality, and abandoning it would allow local hidden variables to explain quantum mechanics. We show here that if an arbitrarily small amount of measurement independence can be guaranteed, then every pure 2-qubit state, except for product states and the maximally entangled state, can produce correlations which cannot be reproduced by measurement dependent local resources. In other words, quantum mechanics resists arbitrary lack of measurement independence, sometimes also called lack of free choice. Besides the immediate implication that the fundamental explanation of nature cannot be limited measurement dependent local correlations, it is of interest that it is possible to make use of the properties of nonlocal correlations even when measurement independence is not guaranteed, which could be of use for the tasks of randomness generation or quantum cryptography.\\

In the following we perform an analysis of the MDL-polytope more complete than the one presented in \cite{Putz14}. In section \ref{polytopes}, we explained that when it comes to the question of membership, i.e. whether a certain point lies within a set, polytopes have the useful property that they can be fully described by a set of linear inequalities called facets. Any point satisfying all the inequalities lies within the set, any point violating any of the inequalities lies outside. The scenario of 2 parties with binary inputs and outputs is dimensionally small enough that we managed to solve the polytope (i.e. find all its facets, cf. section \ref{polytopes}) for fixed values of $\ell$ and $h$, from which we rederived the inequalities as a function of $\ell$ and $h$. While we can easily check that the inequalities are valid for the given range of $\ell$ and $h$ considered, we cannot guarantee that in fact there are no new facets appearing. We do conjecture, however, that this is not the case and that the lists of facets are actually complete. We write the inequalities in the form
\begin{align}
\sum_{abxy}\beta_{ab}^{xy}p(abxy)\leq 0.
\end{align}
Due to symmetries, we can classify the facets into families. Given one member of a family, the others can be found by exchanging the parties ($\beta_{ab}^{xy}\rightarrow\beta_{ba}^{yx}$), flipping Alice's or Bob's input ($\beta_{ab}^{xy}\rightarrow\beta_{ab}^{(x\oplus 1)y}$ or $\beta_{ab}^{xy}\rightarrow\beta_{ab}^{x(y\oplus 1)}$), flipping Alice's or Bob's outcome ($\beta_{ab}^{xy}\rightarrow\beta_{(a\oplus 1)b}^{xy}$ or $\beta_{ab}^{xy}\rightarrow\beta_{a(b\oplus 1)}^{xy}$) or any combination thereof. 

We consider the following three cases:
\begin{itemize}
\item $\ell=h=\frac{1}{4}$: In this case it follows that $p(xy|\lambda)=p(xy)=\frac{1}{4}$, and we are therefore back to full measurement independence. This case thus corresponds to the standard local polytope.
\item $\ell>0$, $h=1-3\ell$: Only a lower bound is imposed, the bound on $h$ follows from (\ref{lhbounds}). For this case, we find $74$ facets, which can be found in table B.1.
\item $\frac{1}{4}<h<\frac{1}{3}$, $\ell=1-3h$: Only an upper bound is imposed, the lower bound $\ell$ follows from (\ref{lhbounds}). We find 93 facets, which are given in table B.2.
\item $h\geq\frac{1}{3}$, $\ell=0$: In this case, measurement dependent local correlations can reproduce all (2,2,2) nonsignaling correlations\footnote{This can be seen by explicitely writing the nonsignaling vertices as a convex combination of MDL vertices. Intuitively, $\ell=0$ means that in each run, there is only one input for which Alice does not know Bob's input and vice-versa. In such a case local correlations can reproduce all nonsignaling correlations.} (cf section \ref{lnsmdl}), and thereby all (2,2,2) quantum correlations. It is therefore not interesting to analyze this case for our purposes. 
\end{itemize}

There are of course other possible ranges of values for $\ell$ and $h$. However it follows from the definition of the MDL-correlations that for $\ell'\leq\ell$ and $h'\geq h$ we have that $\mathcal{MDL}(\ell,h)\subset\mathcal{MDL}(\ell',h')$. As long as if we only wish to prove non-membership for a given $(\ell,h)$ we can thus just check for non-membership of $\mathcal{MDL}(\ell,1-3\ell)$.\\

As already mentioned, all pure partially entangled bipartite states can violate inequality (\ref{goldenineq}) $\forall\ell>0$. They can thus produce correlations which cannot be reproduced by any MDL model with $\ell>0$. Part of the motivation for this complete analysis stems from the fact that the maximally entangled two-qubit state 
\begin{align}
\ket{\Psi}=\frac{1}{\sqrt{2}}\big(\ket{00}+\ket{11})
\end{align}
seems to be singled out. This could have been due to our choice of inequality. However, we can now conduct numerical searches for all the inequalities of the polytope for a fixed value of $\ell$ and $h=1-3\ell$\footnote{Note that while we can only conjecture that the families of facets we found are complete for the whole range of parameters considered, we can verify that for fixed values of the parameters they are indeed complete.} for correlations arising from projective measurements onto the maximally entangled state.

We conducted this numerical search and found that, surprisingly, the maximally entangled state, which leads to the largest violation of the standard locality inequalities in the $(2,2,2)$ scenario, does not seem to allow for a violation of any MDL-inequality for $\ell\leq 0.14$ and $h=1-3\ell$. We conjecture therefore that there exists an MDL-model with $\ell=0.14$ that reproduces the correlations of the maximally entangled 2-qubit state in the case of binary inputs and outputs. For nonlocality experiments and applications taking measurement dependence into account it is therefore recommendable to use partially entangled states.\\

\section{More general scenarios}
\label{secgeneral}
In this section, we consider more general scenarios, meaning scenarios with more than 2 parties or with additional inputs and outputs. Due to the computational complexity, we were not able to find the full set of facets for such scenarios. However, we introduce a procedure to transform any Bell inequality\footnote{Meaning an inequality for the set of local correlations.} into a valid MDL inequality. This allows one to easily use the vast amount of inequalities which have been derived for the local polytope to immediately derive valid inequalities for the MDL polytope. In addition, we present one inequality which is not of this form and which can be violated even for $\ell=0$, something which is impossible in the $(2,2,2)$ case.

\subsection{Transforming any Bell inequality into an MDL inequality}
\label{belltomdl}
Consider a general Bell inequality, holding for all local correlations, i.e. correlations of the form (\ref{local}), given by
\begin{align}
\label{generalbellineq}
\sum_{\{a_i\}_{i=1}^N,\{x_i\}_{i=1}^N}\beta_{\{a_i\}}^{\{x_i\}}p(\{a_i\}_{i=1}^N|\{x_i\}_{i=1}^N)\leq B,
\end{align}
with $\beta_{\{a_i\}}^{\{x_i\}}\in\mathbb{R}$. Then we can derive an MDL inequality in the following way. Let $p_{\{A_i\}_{i=1}^N,\{X_i\}_{i=1}^N}$ be an MDL correlation. Then

\begin{align*}
&\ell\sum_{\beta_{\{a_i\}}^{\{x_i\}}\geq 0}\beta_{\{a_i\}}^{\{x_i\}}p(\{a_i\}_{i=1}^N,\{x_i\}_{i=1}^N)+h\sum_{\beta_{\{a_i\}}^{\{x_i\}}< 0}\beta_{\{a_i\}}^{\{x_i\}}p(\{a_i\}_{i=1}^N,\{x_i\}_{i=1}^N)\\
&=\int\dd\lambda\rho(\lambda)\Big(\ell\sum_{\beta_{\{a_i\}}^{\{x_i\}}\geq 0}\beta_{\{a_i\}}^{\{x_i\}}p(\{x_i\}_{i=1}^N|\lambda) \prod_ip(a_i|x_i\lambda)+h\sum_{\beta_{\{a_i\}}^{\{x_i\}}<0}\beta_{\{a_i\}}^{\{x_i\}}p(\{x_i\}_{i=1}^N|\lambda) \prod_ip(a_i|x_i\lambda)\Big)\\
&\leq \ell h\sum_{\{a_i\}_{i=1}^N,\{x_i\}_{i=1}^N}\beta_{\{a_i\}}^{\{x_i\}}\int\dd\lambda\rho(\lambda)\prod_ip(a_i|x_i\lambda)\\
&\leq \ell h B.
\end{align*}

To get to the second line we simply used the definition of an MDL correlation. The third line follows from $\ell\leq p(\{x_i\}_{i=1}^N|\lambda)\leq h$, using the lower bound for the terms with positive $\beta_{\{a_i\}}^{\{x_i\}}$ and the upper bound for the terms with negative $\beta_{\{a_i\}}^{\{x_i\}}$. Finally the last line follows from the fact that $\int\dd\lambda\rho(\lambda)\prod_ip(a_i|x_i\lambda)$ is by definition a local distribution (\ref{local}) and therefore respects by assumption the Bell inequality (\ref{generalbellineq}).\\

We can use this to generate MDL inequalities from all known Bell inequalities. In fact, inequality (\ref{goldenineq}) could have been derived this way by starting from the Eberhard inequality (\ref{Eberhard}). Since Bell inequalities have been a focus of research for several decades, inequalities have been found in many scenarios and for many purposes. All of these inequalities can, using the technique demonstrated here, directly be transformed into valid MDL inequalities. Even further, since local correlations are nonsignaling (\ref{nspolytope}) and thereby respect equalities of the form
\begin{align}
\label{NScond}
\sum_{a_j}p(\{a_i\}_{i=1}^N|x_1\ldots x_j\ldots x_N)=\sum_{a_j}p(\{a_i\}_{i=1}^N|x_1\ldots x_j'\ldots x_N)\quad \forall x_j,x_j',\{x_i\}_{i\neq j}
\end{align}
one can write Bell inequalities in many different forms equivalent under the nonsignaling constraints. For example, the Eberhard inequality (\ref{Eberhard}) is equivalent to the CHSH inequality
\begin{align}
\label{CHSH}
\sum_{abxy}(-1)^{(a\oplus b)\oplus xy}p(ab|xy)\leq 2
\end{align}
given that the nonsignaling conditions are satisfied. However, since the MDL polytope contains signaling distributions and thereby does not satisfy the nonsignaling conditions (\ref{NScond}), all of these for local correlations equivalent forms of an inequality lead to different valid MDL inequalities by the procedure described above. Specifically the CHSH inequality (\ref{CHSH}) becomes the MDL inequality
\begin{align}
\ell\sum_{(a\oplus b)\oplus xy = 0}p(abxy) - h\sum_{(a\oplus b)\oplus xy = 1}p(abxy)\leq 2\ell h.
\end{align}

Of course, by construction, MDL inequalities derived in this manner can only be useful if $\ell>0$, otherwise they are trivially satisfied by all correlations. In the $(2,2,2)$ scenario this may be fine since the $\ell=0$ case is not of interest anyway. In more general scenarios however, there can be MDL inequalities which are of interest even for $\ell=0$. We mention such an example in the next subsection.\\

There is another way to use Bell inequalities in an MDL scenario: one can simply check what value MDL correlations can achieve for a given Bell inequality and fixed $\ell$ and $h$. Since Bell inequalities are linear, the largest value of a Bell inequality is realized by one of the vertices and it is therefore sufficient to check the values of the Bell inequality for all the MDL vertices. For the case of $h=1-3\ell$ and the CHSH inequality (\ref{CHSH}), we find\footnote{Note that the bound is not actually smaller than the local bound of 2. The switch from conditional probabilities $p(ab|xy)$ to full probabilities $p(abxy)$ involves multiplying by $p(xy)$.}
\begin{align}
\sum_{abxy}(-1)^{(a\oplus b)\oplus xy}p_{MDL}(abxy)\leq 1-2\ell.
\end{align}
This value can even be achieved by a mix of the vertices such that $p(xy)=\frac{1}{4}$, meaning that the inputs appear to be fully random. For this case, the best quantum value is given by~\cite{Tsirelson}
\begin{align}
\sum_{abxy}(-1)^{(a\oplus b)\oplus xy}p_{Q}(abxy)\leq \frac{1}{\sqrt{2}}.
\end{align}
This implies that for $\ell\leq \frac{2-\sqrt{2}}{4}$, quantum mechanical correlations can no longer violate this inequality. It is therefore clearly suboptimal. This shows the usefulness of the MDL polytope when compared to using Bell inequalities for the standard local polytope with an adjusted bound.

\subsection{The (2,n,2) scenario}
One of the major points for the (2,2,2) scenario is that MDL correlations with $\ell=0$ can reproduce all quantum and nonsignaling correlations. The question remained whether $\ell>0$ is a necessary assumption to be able to show quantum measurement dependent nonlocality in general. Adding additional inputs to each party allows us to answer this question in the negative: it is possible to violate MDL inequalities even for $\ell=0$. Consider the (2,n,2) scenario. As in the (2,2,2) scenario, we call the two parties Alice and Bob and refer to their inputs and outputs by $x$ and $y$ and $a$ and $b$ respectively. We can then construct the following inequality, which holds for all MDL correlations $p_{ABXY}$:
\begin{align}
\label{nmineq}
\big(1-(n^2-n+1)h\big)p(0000)-h\sum_{i=1}^{n-1}\big(p(10i0)+p(010i)+p(00ii)\big)\leq 0.
\end{align}
We are using inequality (\ref{goldenineq}) several times between input $0$ and input $i$ for all $i\geq 1$. The factor $(1-(n^2-n+1)h$ makes the inequality trivially satisfied if $h\geq\frac{1}{n^2-n+1}$. If $h<\frac{1}{n^2-n+1}$, then for every $\lambda$ at most $n-2$ inputpairs can be excluded, i.e. have 0 probability due to the normalization of probabilities (\ref{normalization}). We are therefore guaranteed that $\exists i\in\{1\ldots n\}$ such that $p(xy|\lambda)\geq 1-(n^2-n+1)h>0$ for $xy=(0i),(i0),(ii)$. We can then restrict ourselves to these two inputs ($0$ and $i$) and the analysis reduces to the case of $\ell= 1-(n^2-n+1)h>0$ for binary inputs, in which case the inequality corresponds to (\ref{goldenineq}). Every $\lambda$ may have a different $i$ for which this analysis can be done, but the inequality will still hold.\\

This generalized inequality can be violated by quantum mechanics, using the same state and measurements as for inequality (\ref{goldenineq}) and simply doing the same measurement $\forall i\in\{1\ldots n\}$. Of course taking the inputs to be different even though the same measurements are made may be questionable, but it nonetheless shows that $\ell>0$ is not a necessary condition for quantum mechanics to be measurement dependent nonlocal.\\



\subsection{The (N,2,2) scenario}
Let us turn our attention towards the case of $N$ parties, with binary inputs and outputs for each party. Here we will employ the procedure presented in section \ref{belltomdl}. We start with the Bell inequality
\begin{align}
\label{Nbellineq}
p(0\ldots 0|0\ldots 0)-\sum_{j=1}^N p(0\ldots 0\underbrace{1}_j0\ldots 0|0\ldots 0\underbrace{1}_j0\ldots 0) -p(0\ldots 0|1\ldots 1)\leq 0.
\end{align}
Let us briefly show that this inequality holds for all local correlations. Since it is a linear inequality and the local set forms a polytope, its maximal value will be realized by a vertex. The vertices of the local polytope have the properties that they factorize, i.e. $p_{\{A_i\}|\{X_i\}}=\prod_i p_{A_i|X_i}$ and that they are deterministic, i.e. $p(\{a_i\}|\{x_i\})\in \{0,1\}$. To violate the inequality, they therefore have to fulfil
\begin{align*}
\prod_{i=1}^Np_{A_i|X_i}(0|0)&=1\\
p_{A_j|X_j}(1|1)\prod_{i\neq j}p_{A_i|X_i}(0|0)&=0\quad\forall j\in\{1\ldots N\}\\
\prod_{i=1}^Np_{A_i|X_i}(0|1)&=0.
\end{align*}
From the first line it follows that $p_{A_i|X_i}(0|0)=1$ $\forall i$. With this, the second line implies that $p_{A_i|X_i}(1|1)=0$ $\forall i$, which due to the normalization of probabilities implies that $p_{A_i|X_i}(0|1)=1$ $\forall i$. Given this however it follows that $\prod p_{A_i|X_i}(0|1)=1$, which contradicts the third condition. Therefore none of the vertices of the local polytope can violate the linear inequality (\ref{Nbellineq}), implying that no local distribution can.\\

We can now use the procedure shown in section \ref{belltomdl}. We arrive at the MDL inequality
\begin{align}
\ell p(0\ldots 0|0\ldots 0)-h\Big(\sum_{j=1}^N p(0\ldots 0\underbrace{1}_j0\ldots 0|0\ldots 0\underbrace{1}_j0\ldots 0) -p(0\ldots 0|1\ldots 1)\Big)\leq 0.
\end{align}

This inequality can be violated $\forall N$, $\forall \ell>0$ and $\forall h$ by the state
\begin{align}
\ket{\Psi_N}=\frac{1}{\sqrt{N+1}}\big(\ket{0\ldots 01}+\ldots+\ket{10\ldots 0}-\ket{1\ldots 1}\big)
\end{align}
if all $N$ parties perform measurements in the basis $\big\{\frac{\ket{0}+\ket{1}}{\sqrt{2}},\frac{\ket{0}-\ket{1}}{\sqrt{2}}\big\}$ for input 0 and in the basis $\{\ket{0},\ket{1}\}$ for input 1. The left-hand side of inequality (\ref{Nbellineq}) evaluates, for these correlations, to $\frac{(N-1)^2}{(N+1)2^N}$.

\section{Independent sources}
\label{secindsources}
Let us reconsider the MDL-scenario as it is depicted in figure \ref{figmdlscenario}. An intuitive assumption can be seen which we have not imposed so far: any dependence between the inputs is due to the common past $\Lambda$. Mathematically, this means that conditioned on $\Lambda$, the input distribution factorizes:
\begin{align}
\label{indsource}
p_{XY|\Lambda}=p_{X|\Lambda}p_{Y|\Lambda}.
\end{align}
This is a very natural assumption, especially when we consider the common past to be in the hands of an adversary. We call eq. (\ref{indsource}) the \textit{independent source} assumption. For the more general scenario of $N$ parties, it is given by
\begin{align}
p_{\{X_i\}|\Lambda}=\prod_{i=1}^N p_{X_i|\Lambda}.
\end{align}
The measurement dependence assumption is then given by
\begin{align}
\label{mdlis}
\ell_i\leq p(x_i|\lambda)\leq h_i\quad\forall i,x_i,\lambda.
\end{align}
Notice that we allow for different bounds being imposed for the different parties. The set of possible $p_{X_i|\Lambda}$ forms a polytope since only linear inequalities are imposed, and we can therefore use the polytope theorem \ref{polytopethm} to conclude that the resulting set of possible input distributions forms again a polytope. As a consequence, the set of independent source measurement dependent local correlations is also a polytope and can be described by a finite set of linear inequalities. For the case of 2 parties with binary inputs and outputs, we give a list of facets in the appendix, which, as in section \ref{222}, we conjecture to be complete. Note that if the input of party $i$ is binary, it follows that $\ell_i=1-h_i$ due to normalization.\\

It is worth to note that for given $\{\ell_i\}_i$ and $\{h_i\}_i$, the set of independent sources MDL distributions is included in the set of MDL correlations with $\ell=\prod_i\ell_i$ and $h=\prod_i h_i$. In the case of 2 parties with binary inputs and outputs, all pure partially entangled quantum states thus allow for correlations which cannot be reproduced by independent sources MDL distributions if $\ell_i>0$ $\forall i$. We investigated whether with this additional assumption 2-qubit maximally entangled states could produce such correlations as well. However our numerical optimizations yielded no such results, and we conjecture that for some critical $\ell_i>0$ independent sources MDL resources can reproduce all binary input and outcome correlations of the maximally entangled 2-qubit state.

\section{Measurement dependence and signaling}
We noted at several points that measurement dependent correlations are not nonsignaling. By that we mean that a measurement dependent correlation does not necessarily satisfy the nonsignaling conditions\footnote{For simplicity, we only talk about the bipartite scenario here.}
\begin{align}
p(a|xy)=p(a|xy')\quad\forall a,x,y,y',
\end{align}
and similarly for Bob. If these conditions are not satisfied, then it is possible for one party to gain information about the input of the other by only looking at their own input and output. To see that measurement dependent local correlations do not necessarily satisfy the nonsignaling conditions, it suffices to consider the definition of the marginals:
\begin{align*}
p(a|xy)&=\frac{\sum_bp(abxy)}{p(xy)}\\
&\int\dd\lambda\rho(\lambda)\frac{p(xy|\lambda)}{p(xy)}p(a|x\lambda),
\end{align*}
which in general is not independent of $y$.
Usually, one would say that this allows for communication that is faster than light or without physical support. We would like to stress that this conclusion does not hold here. In fact, reconsidering the definition of measurement dependent local correlations, it becomes clear that by definition, if $\lambda$ is known, then the nonsignaling conditions are satisfied, i.e.
\begin{align}
p(a|xy\lambda)=p(a|xy'\lambda)\quad\forall a,x,y,y',\lambda.
\end{align}
It is thus evident that no faster than light communication is happening here. Instead it can simply be seen as the common source $\Lambda$ influencing and potentially correlating the four random variables $A, B, X$ and $Y$. This is an example of a case where the illusion of faster than light communication arises due to ignorance, in this case of the local hidden variable $\Lambda$. The same conclusion holds for the case of independent sources described in section \ref{secindsources}.\\

If one is certain however that the nonsignaling conditions should be fulfilled for some observed correlations, then our characterization of the MDL polytope can still easily be used for the analysis. To do so, note that adding linear equality constraints to the constraints of a polytope results in a new polytope. Geometrically, it corresponds to cutting a polytope with the hyperplane defined by the additional linear equality constraints. Given that the facets, i.e. the hyperplanes describing the original polytope, are known, finding the description of the new polytope corresponds to computing the intersection of the facets with the constraint-hyperplane: a case of variable elimination. In this way, one gets the facets of the new polytope with additional superfluous hyperplanes, which can be eliminated by a linear program.

We performed this operation to get the nonsignaling measurement dependent local polytope for the case of bipartite correlations with binary inputs and outputs and $\frac{1}{4}< h<\frac{1}{3}$, $\ell=1-3h$ in our previous work~\cite{Putz14}. The total number of facets reduced drastically from 93 to 8.

\section{Limited detection locality}
\label{LDL}
Let us mention another way in which local resources can reproduce correlations that are seemingly nonlocal. Consider an experiment in which the particles used to establish the correlations can be lost. One may, in such a case, want to simply ignore the cases in which such a loss occurred, i.e. postselect on having a detection for all the involved parties. This however opens up the detection loophole. What could happen in fact is that the particles have the hidden common strategy to simply not give a detection if they do not like the input they see once they arrive in the respective measurement apparatus. If these cases are simply ignored, then this corresponds to the experimenters letting the particles choose which input they reply to, which leads us back to the problem of measurement dependence.\\

The usual way to deal with this issue is to either have a sufficiently large overall efficiency, such that this cheating strategy can be excluded, or to simply ignore it. Given that the necessary detection efficiencies are usually very high, this seems very unsatisfactory. We recently introduced the assumption of \textit{limited detection efficiency}\cite{Putz15}. It corresponds to assuming that in each run the detection probability cannot be fully influenced by the common hidden strategy. In other words, if we denote by $\varnothing$ a nondetection event, then we assume that
\begin{align}
\eta_{min}\leq p(A_i\neq\varnothing|x_i\lambda)\leq\eta_{max}
\end{align}
for a fixed $[\eta_{min},\eta_{max}]\subsetneqq [0,1]$. Amongst the interesting results is the fact that the set of correlations realizable by postselected local distributions with this assumption forms a polytope. We do not discuss these results here and instead refer to \cite{Putz15}.\\

For the purposes of this paper, it is however interesting that the connection between limited detection and measurement dependence, as intuitively described above, can be made mathematically explicit. In fact, it turns out that any set of postselected limited detection local distributions can also be realized by measurement dependent local distributions and even further, we can analyze them both together. This is captured by the following theorem:
\begin{thm}
Let 
\begin{align*}\mathcal{P}(\ell,h,\eta_{min},\eta_{max})=\Big\{p :& p(\{a_i\}_{i=1}^N,\{x_i\}_{i=1}^N)=\int\dd\lambda\rho(\lambda)p(\{x_i\}_{i=1}^N|\lambda)\prod_{i=1}^N p(a_i|x_i\lambda)\\
&\ell\leq p(\{x_i\}_{i=1}^N|\lambda)\leq h\quad\forall x,y,lambda\\
&\eta_{min}\leq p(a_i\neq\varnothing|x_i\lambda)\leq\eta_{max}\quad\forall i,a_i,x_i,\lambda\Big\}
\end{align*}
and
\begin{align*}
\mathcal{PS}(\ell,h,\eta_{min},\eta_{max})=\Big\{q : \exists p\in\mathcal{P}(\ell,h,\eta_{min},\eta_{max})\text{ s.t. }q_{\{A_i\},\{X_i\}}=p_{\{A_i\},\{X_i\}|A_i\in\{1\ldots m_i\}\forall i}\Big\}
\end{align*}
Then $\mathcal{PS}(\ell,h,\eta_{min},\eta_{max})\subset \mathcal{MDL}\Big(\ell\big(\frac{\eta_{min}}{\eta_{max}}\big)^2,h\big(\frac{\eta_{max}}{\eta_{min}}\big)^2\Big)$.
\end{thm}

The proof of this theorem can be found in \cite{Putz15}. If we want to guarantee that a certain correlation cannot be replicated by measurement dependent and limited detection local resources, we can instead focus only on measurement dependent resources with adjusted parameters $\ell'=\ell\big(\frac{\eta_{min}}{\eta_{max}}\big)^2$ and $h'=h\big(\frac{\eta_{max}}{\eta_{min}}\big)^2$. Thereby all the results introduced in this work can be applied.

\section{Conclusion}
Bell-locality, one of the most fascinating features of quantum mechanics and the core concept behind the framework of device independent quantum information processing, includes the untestable assumption of measurement independence. We relaxed this assumption and found that the correlations we get with our parameterized relaxation, which we refer to as measurement dependent local correlations, form a convex polytope, a geometric structure that can be fully characterized by a finite number of linear inequalities. We found the complete set of inequalities for the case of bipartite correlations with binary inputs and outputs. Using one of these inequalities, we found that, surprisingly, all bipartite pure states except for the seperable states and the maximally entangled state can exhibit nonlocality for arbitrarily large measurement dependence, i.e. an arbitrarily small amount of free choice. This is again an example of nonlocality and entanglement being not in one-to-one correspondence~\cite{Brunner05}.

Furthermore, we presented a method allowing us to transform any Bell inequality into a valid MDL inequality, allowing us to use all of the already established Bell inequalities for our purposes. Using this we gave a family of MDL inequalities for $N$ parties with binary inputs and outputs. We also presented a family of bipartite inequalities for arbitrary number of inputs, with which we could show that our measurement dependence assumption can be relaxed even further while still allowing for quantum violations.

We presented the additional assumption of independent sources, which in the given framework comes very natural. We showed that this case can be fully described using linear inequalities as well and gave all the inequalities for the bipartite case with binary inputs and outputs.

Finally, we considered the more realistic scenario of a lossy experiment or implementation. We introduced the concept of limited detection locality. Interestingly, there exists a direct link between this limited detection scenario and the measurement dependence scenario.

Several open questions remain, which we exhibit here for potential future research. First, the definition of measurement dependence leads to considerations of randomness amplification or extraction. It would be of interest to see if the measurement dependent local inequalities presented here can be used to certify randomness, potentially in the bipartite case with binary inputs and outputs.

Second, it would be of interest to see whether the set of measurement dependent local distributions is closed under so-called wirings, i.e. when the outputs of one setup are used as the inputs for another. 

Finally, the current framework assumes that the probability distribution describing a run of the experiment is known. In a real experiment, this is accomplished by checking frequencies over multiple runs. Our results apply straightforwardly if we assume that we do not need to take memory effects into account, i.e. that the runs are independent and identically distributed. Taking such memory effects into account is a topic we are currently working on and will display in future publications.

\section{Acknowledgments}
We acknowledge Tomer Barnea and Denis Rosset for disccusions and financial support of the European projects CHIST-ERA DIQIP and SIQS.

\bibliography{MDLFullLength}

\newpage

\appendix

\section{Proof of theorem 1}
\begin{thm*}
Let $\mathcal{P}\subset\mathbb{R}^{m'n}$ and $\mathcal{Q}\subset\mathbb{R}^{mn}$ be two polytopes. Let the components of the vectors of $\mathcal{P}$ be labelled by $p(k,l)$ and the ones of $\mathcal{Q}$ by $q(k,l')$ where $k=1\ldots n$, $l=1\ldots m$ and $l'=1\ldots m'$. Let
\begin{align*}
\mathcal{R}=\big\{r\in\mathbb{R}^{mm'n} : &\exists\Lambda\text{ measureable set},\\
&\exists\rho:\Lambda\rightarrow [0,1]\text{ with }\rho(\lambda)\geq 0\forall\lambda\in\Lambda,\int_\Lambda\dd\lambda\rho(\lambda)=1\\
&\exists \{p_\lambda\}_{\lambda\in\Lambda}\subset\mathcal{P}, \quad\exists \{q_{\lambda}\}_{\lambda\in\Lambda}\subset\mathcal{Q} \text{ s.t.}\\
&r(k,l,l')=\int\dd\lambda\rho(\lambda)p_\lambda(k,l)q_{\lambda}(k,l')\quad \forall k\in\{1\ldots n\},l\in\{1\ldots m\},l'\in\{1\ldots m'\}\big\}.
\end{align*}
Let $\mathcal{V_P}$ and $\mathcal{V_Q}$ be the sets of vertices of $\mathcal{P}$ and $\mathcal{Q}$ respectively.\\
Then $\mathcal{R}$ is a polytope and its set of vertices $\mathcal{V_R}$ fulfils
\begin{align*}
\mathcal{V_R}\subset\big\{v_\mathcal{R}^{ij} : v_\mathcal{R}^{ij}(k,l,l')=v_\mathcal{P}^i(k,l)v_\mathcal{Q}^j(k,l')\quad i\in\{1\ldots |\mathcal{V_P}|\},j\in\{1\ldots |\mathcal{V_Q}|\}\big\}.
\end{align*}
\end{thm*}

\textbf{Proof:} To prove the theorem, we show the following 3 steps:
\begin{enumerate}
\item $\mathcal{R}$ is convex.
\item $\mathcal{V_R}\subset\mathcal{R}$
\item $\mathcal{R}\subset\text{Conv}(\mathcal{V_R})$, where $\text{Conv}$ denotes the convex hull.
\end{enumerate}

\textit{Step 1: }Let $r_1,r_2\in\mathcal{R}$, $\alpha\geq 0$ and $r=\alpha r_1+(1-\alpha)r_2$. Then
\begin{align*}
r(k,l,l')&=\alpha r_1(k,l,l')+(1-\alpha)r_2(k,l,l')\\
&=\int_{\Lambda_1}\dd\lambda_1\alpha\rho_1(\lambda_1)p^1_{\lambda_1}(k,l)q^1_{\lambda_1}(k,l')+\int_{\Lambda_2}\dd\lambda_2(1-\alpha)\rho_2(\lambda_2)p^2_{\lambda_2}(k,l)q^2_{\lambda_2}(k,l')\\
&=\int_\Lambda\dd\lambda\rho(\lambda)p_\lambda(k,l)q_\lambda(k,l'),
\end{align*}
where we define $\Lambda=\Lambda_1\times\Lambda_2\times\{1,2\}$ and $\rho : \Lambda=\Lambda_1\times\Lambda_2\times\{1,2\}\rightarrow \mathbb{R}^+$ with
\begin{align*}
\int_{\Lambda_2}\dd\lambda_2\rho(\lambda_1,\lambda_2,1)&=\alpha\rho_1(\lambda_1)\\
\int_{\Lambda_1}\dd\lambda_1\rho(\lambda_1,\lambda_2,2)&=(1-\alpha)\rho_2(\lambda_2)\\
p_{(\lambda_1,\lambda_2,n)}&=p^n_{\lambda_n}\\
q_{(\lambda_1,\lambda_2,n)}&=q^n_{\lambda_n}\\
\end{align*}
Note that $\int_\Lambda\dd\lambda\rho(\lambda)=1$, $p_\lambda\in\mathcal{P}$, $q_\lambda\in\mathcal{Q}$ $\forall\lambda\in\Lambda$. We therefore conclude that $r\in\mathcal{R}$ and thus that $\mathcal{R}$ is convex.\\

\textit{Step 2: }Since $\mathcal{V_P}\subset\mathcal{P}$ and $\mathcal{V_Q}\subset\mathcal{Q}$, we conclude that by definition $\mathcal{V_R}\subset\mathcal{R}$.\\

\textit{Step 3: }Let $r\in\mathcal{R}$. Then
\begin{align*}
r(k,l,l')&=\int_\Lambda\dd\lambda\rho(\lambda)p_\lambda(k,l)q_\lambda(k,l')\\
&=\int_\Lambda\dd\lambda\rho(\lambda)\sum_i\alpha^i_\lambda v_\mathcal{P}^i\sum_j\beta^j_\lambda v_\mathcal{Q}^j\\
&=\sum_{i,j}\Big(\int_\Lambda\dd\lambda\rho(\lambda)\alpha^i_\lambda\beta^j_\lambda\Big)v_\mathcal{P}^iv_\mathcal{Q}^j\\
&=\sum_{i,j}\gamma^{ij}v_\mathcal{R}^{ij}.
\end{align*}
In the first line we use the definition of $\mathcal{R}$, in the second line we use that $\mathcal{P}$ and $\mathcal{Q}$ are polytopes and thus that we can express their elements as convex combinations of their vertices. Note that therefore $\sum_i\alpha_\lambda^i=1$ $\forall\lambda$, $\sum_j\beta_\lambda^j=1$ $\forall\lambda$ and $\alpha_\lambda^i,\beta_\lambda^j\geq 0$ $\forall i,j,\lambda$. In the last line we defined $\gamma^{ij}=\int_\Lambda\dd\lambda\rho(\lambda)\alpha^i_\lambda\beta^j_\lambda$. Note that $\gamma^{ij}\geq 0$ and $\sum_{i,j}\gamma^{ij}=1$. 

We can thus write any element of $\mathcal{R}$ as a convex combination of elements of $\mathcal{V_R}$ and therefore $\mathcal{R}\subset\text{Conv}(\mathcal{V_R})$.\\

This concludes the proof.

\section{Full list of facets for the (2,2,2) scenario and some ranges of l and h}
In this section we give the full list of facets for the cases of $0<\ell<\frac{1}{4}$, $h=1-3\ell$, corresponding to the case where only a nontrivial lower-bound is imposed, as well as $\frac{1}{4}<h<\frac{1}{3}$, $\ell=1-3h$, where only an upper bound is imposed. The tables contain the $\beta_{ab}^{xy}$ such that $\sum_{abxy}\beta_{ab}^{xy}p(abxy)\leq 0$ for all MDL($\ell,h$)-correlations $p_{ABXY}$.

}

\caption{Facets of the MDL($\ell,h$) polytope for $\frac{1}{4}<h<\frac{1}{3}$ and $\ell=1-3h$. The facets are given by $\sum_{abxy}\beta_{ab}^{xy}p(abxy)\leq 0$.}

\end{sidewaystable}

\section{Full list of facets for independent sources measurement dependent local distributions}
Here we give a list of hyperplanes which we conjecture to be the complete list of facets for the case of independent sources.

\begin{sidewaystable}
\resizebox*{\textwidth}{!}{\begin{tabular}{c|c|c|c|c|c|c|c|c|c|c|c|c|c|c|c}
$P(0000)$&$P(1000)$&$P(0100)$&$P(1100)$&$P(0010)$&$P(1010)$&$P(0110)$&$P(1110)$&$P(0001)$&$P(1001)$&$P(0101)$&$P(1101)$&$P(0011)$&$P(1011)$&$P(0111)$&$P(1111)$\\
\hline

 $\frac{(2 h_y-1) h_x^2+2 h_y^2 h_x-h_y^2}{(h_x+h_y-1) (-h_y+h_x (2h_y-1)+1)}$&$1$&$1$&$1$&$\frac{h_y}{h_y-1}$&$\frac{(2 h_x-1) h_y^4+2 (h_x-1)^2 (h_y^3-5 h_y^2+4 h_y-1)}{(h_y-1) h_y (h_x+h_y-1) (-h_y+h_x (2h_y-1)+1)}$&$\frac{h_y}{h_y-1}$&$\frac{h_y-1}{h_y}$&$\frac{h_x}{h_x-1}$&$\frac{h_x}{h_x-1}$&$\frac{2 h_y+(h_x-1) \left((h_x-1) (2 h_x-1) h_y^2+2 h_x ((h_x-1) h_x+4) h_y+h_x\left(-h_x^2+h_x-4\right)\right)-1}{(h_x-1) h_x (h_x+h_y-1) (-h_y+h_x (2h_y-1)+1)}$&$\frac{h_x-1}{h_x}$&$\frac{h_x h_y \left((2 h_y-1) h_x^2+2 ((h_y-4) h_y+2)h_x-(h_y-4) h_y-2\right)}{(h_x-1) (h_y-1) (h_x+h_y-1) (-h_y+h_x (2h_y-1)+1)}$&$\frac{h_x (h_y-1)}{(h_x-1) h_y}$&$\frac{(h_x-1) h_y}{h_x(h_y-1)}$&$\frac{h_x^2 (h_y-1)^2-2 h_x h_y^2+h_y^2}{(h_x-1) h_x (h_y-1)h_y} $ \\

 $\frac{(h_x-1)^2 h_y^2}{h_x^2 (h_y-1)^2}$&$1-\frac{(2 h_x-1) h_y^2}{h_x^2(h_y-1)^2}$&$\frac{(h_x-1)^2 h_y^2}{h_x^2 (h_y-1)^2}$&$1$&$\frac{h_y}{h_y-1}$&$\frac{(h_x-1)^2h_y}{h_x^2 (h_y-1)}$&$\frac{h_y}{h_y-1}$&$\frac{h_y}{h_y-1}$&$\frac{(h_x-1)h_y^2}{h_x (h_y-1)^2}$&$\frac{h_x-1}{h_x}$&$-\frac{h_y^2 \left((1-2 h_x)^2 h_y^2+((h_x-1) h_x ((h_x-1) h_x+4)+1) (1-2h_y)\right)}{(h_x-1) h_x^3 (h_y-1)^2 (2 h_y-1)}$&$\frac{(h_y-1)^2 (2 h_y-1) h_x^4+2 (1-2h_y)^2 h_x^3-\left(4 h_y \left(h_y^3+h_y-1\right)+1\right) h_x^2+4 h_y^4h_x-h_y^4}{(h_x-1) h_x^3 (h_y-1)^2 (2 h_y-1)}$&$\frac{(h_x-1) h_y}{h_x
   (h_y-1)}$&$\frac{(h_x-1) h_y}{h_x (h_y-1)}$&$-\frac{h_y \left(-2 h_y h_x^4+h_x^4+(1-2h_x)^2 h_y^2\right)}{(h_x-1) h_x^3 (h_y-1) (2 h_y-1)}$&$-\frac{h_y \left((h_x-1)^2h_x^2-2 (h_x-1)^2 h_y h_x^2+(1-2 h_x)^2 h_y^2\right)}{(h_x-1) h_x^3 (h_y-1) (2h_y-1)} $ \\
\end{tabular}}
\caption{Families of facets for independent sources. The equation of the hyperplane is given by $\sum_{abxy}\beta_{ab}^{xy}p(abxy)=0$.}
\end{sidewaystable}

\end{document}